\newif\ifmulticol	\multicoltrue
\newif\ifshowgit	\showgittrue		% switches footer on/off
\newif\ifgitlocal	\gitlocaltrue		% use local file gitHeadLocal.gin
\newif\ifbiblatex	\biblatexfalse		% defaults to bibtex if false
\newif\ifbibnum		\bibnumtrue			% num => superscripts, otherwise auth date
\newif\ifbibsort	\bibsortfalse		% biblatex num sort in order of occurrence
\newif\iflineno		\linenofalse
\newif\iftoc		\tocfalse

\newif\iflucida		\lucidafalse
\newif\ifcm			\cmfalse
\newif\ifcharter	\chartertrue		% use for arXiv, requires xelatex
\newif\ifcharterotf	\charterotffalse	% requires xelatex

\newcommand*{\secnumdepth}{0}			% sets secnumdepth in \mymaketitle
\newcommand*{\tocdepth}{2}				% sets tocdepth in \mymaketitle
\newcommand*{\reftitle}{References}		% title for references section
\newcommand*{\mytitleskip}{-0.2in}		% change skip after title

\newcommand*{\mydocfontsize}{\ifcharter11pt\else10pt\fi}
\newcommand*{\setcol}{\ifmulticol twocolumn\else onecolumn\fi}

\documentclass[\mydocfontsize,\setcol]{article}

\newcommand*{\pdfauthor}{Steven A.~Frank}
\newcommand*{\pdftitle}{Causal attribution by the chain rule: unifying natural selection, learning, economics, and other disciplines}

\input ftns.sty
\newcommand{\bx}{\bmr{x}}
\newcommand{\bb}{\bmr{b}}
\newcommand{\bz}{\bmr{z}}
\newcommand{\bp}{\bmr{p}}
\newcommand{\bq}{\bmr{q}}
\newcommand{\bbx}{\bmr{x}}
\newcommand{\xbar}{\bar{x}}
\newcommand{\bxbar}{\bar{\bbx}}
\newcommand{\wbar}{\bar{w}}
\newcommand{\zbar}{\bar{z}}

\newcommand*{\Ga}{\alpha}

\newcommand*{\GD}{\Delta}
\newcommand*{\Ge}{\epsilon}

\usepackage{bm}
\newcommand{\bmr}[1]{\bm{\mathrm{#1}}}

\DeclarePairedDelimiter\abs{\lvert}{\rvert}
\DeclarePairedDelimiter\norm{\lVert}{\rVert}
\DeclarePairedDelimiter\angb{\langle}{\rangle}
\DeclarePairedDelimiter\lrb{\lbrack}{\rbrack}
\DeclarePairedDelimiter\lr{\lparen}{\rparen}
\DeclarePairedDelimiter\lrbr{\lbrace}{\rbrace}

\makeatletter
\let\oldabs\abs \def\abs{\@ifstar{\oldabs}{\oldabs*}}
\let\oldnorm\norm \def\norm{\@ifstar{\oldnorm}{\oldnorm*}}
\let\oldangb\angb \def\angb{\@ifstar{\oldangb}{\oldangb*}}
\let\oldlrb\lrb \def\lrb{\@ifstar{\oldlrb}{\oldlrb*}}
\let\oldlr\lr \def\lr{\@ifstar{\oldlr}{\oldlr*}}
\let\oldlrbr\lrbr \def\lrbr{\@ifstar{\oldlrbr}{\oldlrbr*}}
\makeatother

\newcommand*{\dd}{\textrm{d}}
\newcommand*{\Eq}[1]{eqn~\ref{eq:#1}}

\newcommand*{\boldrule}{\hrule height 1.2pt}
\newcommand*{\noterule}{\medskip\boldrule\medskip}	% for notes

\hypersetup{linkcolor={Maroon4}, urlcolor={Maroon4}}

\newcommand{\reviset}[1]{{\color{black}#1}}

\newcommand*{\mytitle}{\pdftitle}

\newcommand*{\myauthor}{Steven A.\ Frank}
\newcommand*{\myaddress}{Department of Ecology and Evolutionary Biology, University of California, Irvine, CA 92697--2525, USA}
\newcommand*{\mycontact}{\href{https://stevefrank.org}{https://stevefrank.org}}

\newcommand*{\mykeywords}{Fisher's fundamental theorem of natural selection, Price equation, Oaxaca-Blinder decomposition, counterfactual causal analysis, machine learning, back propagation}
\setabstract{0.0}{\iftoc-2.0\else0.0\fi}{%
Analysis often splits change into components. For example, how much of the observed variance is caused by genes or environment? In many cases, the split is ultimately made by the logic of the chain rule, which divides the difference of a product into two terms. Each term quantifies the partial difference associated with change in one component while holding the other component constant. The chain rule is of course widely known. However, this article argues that its deep fundamental role often goes unrecognized. The article shows how simply the basic chain rule unifies Fisher's fundamental theorem of natural selection, the Price equation description of evolutionary change, the Oaxaca-Blinder decomposition of wage differences in economics, the Kitagawa decomposition of mortality differences in demography, many expressions of thermodynamics, and most strikingly back propagation, the core optimization method of modern machine learning and artificial intelligence. The success in creating good designs and finding good solutions in both natural selection and artificial intelligence depends on how the chain rule propagates causes from instances of success or failure back to the underlying genes or parameters of the system. The mathematical analysis presented here shows that, for finite differences, the product rule form of the chain rule yields a basic decomposition of change into two components of a regression equation. That regression decomposition is purely a description of change with no explicit causal meaning. However, simple additional assumptions lead naturally to the modern counterfactual analysis of causality. From that perspective, we can easily understand the causal interpretation that Fisher gave to his fundamental theorem, and we can see the same causal structure in the Oaxaca-Blinder decomposition of economics and in causal analyses across many disciplines.
}

\begin{document}

\mymaketitle

%% 1st parm is skip on left column at start of TOC, 2nd param is skip after TOC
\iftoc\mytoc{-24pt}{\newpage}\fi

\reviset{
\section{Introduction}

We often split observed variation into components. For example, how much of the change in a trait is caused by natural selection, and how much is caused by a change in the environment? This article shows that a simple common method often underlies such causal decomposition.

The general understanding of method leads to a broad conceptual unification of how one studies cause. And it leads to insight about the common basis by which natural selection and the complex computational algorithms of modern artificial intelligence gain information. Along the way, we also see the common basis of important decompositions for variation and causal analysis in other disciplines, including the famous Oaxaca-Blinder decomposition of economic analysis \autocite{blinder73wage,oaxaca73male-female,fortin11decomposition}.

The mathematical foundation arises from the chain rule, which divides the difference of a product into two terms. Each term quantifies the partial difference caused by change in one factor while holding the other factor constant.  In standard calculus, the application of the chain rule to a product is called the product rule. This simple identity turns out to be the common algebraic foundation for a remarkably wide range of analytical methods across the sciences.

An early example is Fisher's fundamental theorem of natural selection \autocite{fisher30the-genetical,fisher41average,fisher58the-genetical}. Fisher regressed the fitness of individuals on their genetic composition. He then split the total change in mean fitness into two terms. The first ascribed gene frequency changes to natural selection while holding constant the partial regression coefficients of individual genes on fitness. The second term ascribed regression coefficient changes to changes in the genetic context and the environment \autocite{price72fishers,frank97the-price}.

This article shows that Fisher's regression decomposition follows simply from the product rule for finite differences, which is the common name for the chain rule applied to a product when changes are not infinitesimally small.

My concern is not with the value of Fisher's theorem as a statement about mean fitness in populations. Rather, this article sets Fisher's theorem in the broader context of methods to decompose change into parts and to consider the causal interpretation of the components. These methods clarify the history of Fisher's theorem. More importantly, the study of change and the analysis of cause are central goals in all of the sciences.

To develop that point, this article provides a very general mathematical derivation of the regression decomposition into a set of direct factors and a set of contextual factors. That decomposition arises simply from the product rule for finite differences. Fisher's theorem is used as an illustrative example.

Fisher argued that the first term of his regression decomposition isolates the direct force of natural selection from the broader complexity of evolutionary change, just as one might isolate gravity as a force from the actual trajectory of a body that depends on friction, constraints, and initial conditions. That causal assignment of credit for change to a specific factor while holding other factors constant is the logic of the modern counterfactual approach to causality \autocite{holland86statistics,pearl09causality}, which Fisher anticipated by decades.

Price and Ewens showed that Fisher's decomposition is an attempt at causal attribution \autocite{price72fishers,ewens89an-interpretation}. But they both disputed his peculiar definitions and causal interpretation of natural selection. For this article, the primary point is Fisher's underlying setup and application of counterfactual causal analysis. Causal interpretation almost always has a subjective component, and it is Fisher's subjective choice for causal interpretation that many authors do not accept.

\section{Contributions of this work}
 
This article makes three contributions. First, by deriving the minimal algebraic structure, I show that Fisher's theorem, the Oaxaca-Blinder decomposition of economics \autocite{fortin11decomposition}, the Kitagawa decomposition of demography \autocite{kitagawa55components}, the Price equation \autocite{price72extension,frank12naturalb}, and related results in thermodynamics all reduce to the product rule for finite differences, often applied to a regression equation.
 
Second, I show how the essence of modern counterfactual causal analysis arises in a simple and natural way from the chain rule. In particular, the chain rule assigns causal credit by tracing how a change in each component propagates through a system to affect the outcome. The basic regression decomposition that follows from the finite difference product rule form of the chain rule is purely a description of change with no explicit causal meaning. However, a simple interpretive perspective leads naturally to the modern counterfactual analysis of causality.
 
Third, the chain rule is how machine learning uses observed successes and failures to attribute cause to model parameters. Causal attribution provides parameter updates to improve performance \autocite{goodfellow16deep}. Similarly, natural selection finds good designs by implicitly back propagating reproductive success through a chain-rule-like calculation that attributes cause to particular genes, leading to updated gene frequencies. In both cases, causal attribution is hindered by nonlinear interactions. But things typically work well enough to create the seeming miracles of natural and artificial life. I unified natural selection with other types of learning in a related article \autocite{frank25the-price}.
}

\section{Product rule for finite differences}

Finite differences have the form $\GD y = y'-y$. The prime denotes the value of the variable in a changed context. Let $z=bx$ be the product of $b$ and $x$. Then we can write the finite difference for $z$ as
\begin{equation*}
  \GD z=\GD\lr{bx}=b'x'-bx=\lr{b+\GD b}\lr{x+\GD x}-bx,
\end{equation*}
which simplifies to the product rule for finite differences
\begin{equation*}
  \GD z=b\GD x+x'\GD b.
\end{equation*}
On the right side, the first product term is the change in $x$ holding $b$ constant, and the second product term is the change in $b$ multiplied by $x'$, the value of $x$ in the changed context. This is a particular type of chain rule expansion.

We often need the difference of sums. Let
\begin{equation*}
  z=\sum_{i=0}^n b_ix_i=\bb\cdot\bx,
\end{equation*}
in which bold variables are vectors, and the dot denotes the dot product of two vectors, which as shown is the sum of the element-wise products of the vectors. Then the product rule for finite differences extends to
\begin{equation}\label{eq:finiteD}
  \GD z=\bb\cdot\GD\bx \mskip5mu+\mskip5mu \bx'\cdot\GD\bb.
\end{equation}
If the differences are small, we write the difference operator as a differential, $\GD\rightarrow\dd$. Then, recalling that $x'=x+\GD x$ and replacing differences by differentials, we obtain
\begin{equation*}
  \dd z=\bb\cdot\dd\bx \mskip5mu+\mskip5mu \bx\cdot\dd\bb
  			\mskip5mu+\mskip5mu \dd\bx\cdot\dd\bb.
\end{equation*}
For small differences, the final term is the second order product of two small values, which is negligible relative to the first order smallness of the prior terms. Thus, in the limit we drop the final term and get the product rule for differentials
\begin{equation*}
  \dd z=\bb\cdot\dd\bx \mskip5mu+\mskip5mu \bx\cdot\dd\bb.
\end{equation*}
However, many realistic analyses apply to finite differences with measurements at two separated times. So, the less known product rule for finite differences in \Eq{finiteD} is often a useful expression, as shown in the next section.

\section{Difference in a regression model}
 
Write the value of our focal variable as a regression equation
\begin{equation*}
  z_j=b_0 x_{0j}+\sum_{i=1}^n b_i x_{ij} + \Ge_j
\end{equation*}
Here, we have a population of $z$ values indexed by $j$. Each instance of $z_j$ depends on the predictors, $x_{ij}$, with each predictor weighted by its associated partial regression coefficient, $b_i$. The term $\Ge_j$ is the residual or error.

In the intercept term, $b_0 x_{0j}$, going forward we set $x_{0j}=1$ for all $j$, so that $b_0$ is the intercept. This allows us to write the regression more compactly in a way that matches the notation of other key steps in this article as
\begin{equation}\label{eq:regZ}
  z_j=\sum_{i=0}^n b_i x_{ij} + \Ge_j = \bb\cdot\bx_j+\Ge_j.
\end{equation}
Each instance $j$ depends on the vector of predictors, $\bx_j$, with each predictor weighted by its associated partial regression coefficient in $\bb$. Subsequent sums for $i$ also run over $i=0,1,\dots,n$.

In the underlying population of $z$ values, suppose each value $z_j$ occurs with frequency $q_j$. Then the mean value of $z$ is
\begin{align*}
	\zbar = \sum_j q_j z_j=\bq\cdot\bz 
					&= \sum_i b_i\sum_jq_jx_{ij}\\
					&= \sum_i b_i\xbar_i\\
					&= \bb\cdot\bxbar.
\end{align*}
\reviset{Here, by standard frequency-weighted least squares for regression, the average residual is zero, $\sum_j q_j\Ge_j=0$ \autocite{rao73linear,searle71linear,seber03linear}.}

For the regression of $z$, we want the difference between the current mean value and the mean value in a changed context, $\GD\zbar=\zbar'-\zbar$, which we obtain by applying the general expression for finite differences in \Eq{finiteD}, yielding
\begin{equation}\label{eq:regressD}
  \GD\zbar=\GD\lr{\bb\cdot\bxbar}=\bb\cdot\GD\bxbar \mskip5mu+\mskip5mu \bxbar'\cdot\GD\bb.
\end{equation}
It is of course possible to interpret particular predictors as interactions between causal factors. But the point here concerns the general structure of the problem rather than the specific causal model.

\section{Fisher's fundamental theorem}

Fisher's theorem is simply the application of the general expression for the difference in the mean value of a regression variable, given in \Eq{regressD}. Let the variable of interest be fitness, denoted $w\equiv z$, and the difference in mean fitness be $\GD\wbar\equiv\GD\zbar$.

For simplicity, assume a haploid genetic model, in which the multiple loci in an individual are denoted by $x_{ij}$ for the $i=1,\dots,n$ loci in the $j$th individual, with $x_{0j}=1$ as the indicator variable for the intercept. Suppose that at each locus there is an allele that takes on values of either $0$ or $1$. More complex genetics can be handled within similar regression expressions, but the added notation does not provide additional insight \autocite{frank97the-price}.

With those assumptions, the average value of a predictor in the regression is the frequency of the allele with value $1$, or in standard genetic language, $p_i=\xbar_i$ is the gene frequency at the $i$th locus. For the intercept, $p_0=1$.

Now the change in mean fitness follows immediately from \Eq{regressD} as
\begin{equation}\label{eq:wD}
  \GD\wbar = \GD\lr{\bb\cdot\bp}=\bb\cdot\GD\bp \mskip5mu+\mskip5mu \bp'\cdot\GD\bb.
\end{equation}
This expression describes the total change exactly. Fisher emphasized that the first term focuses on the change in gene frequencies caused by natural selection, $\GD\bp$, while holding constant the average effect of each gene on fitness, given by the partial regression coefficients, $\bb$.

\subsection{Main result}

If, without loss of generality, we assume that in the initial population $\wbar=1$, and, following Fisher, define the marginal fitness of allele $i$ as $w_i=p_i'/p_i$, then
\begin{equation*}
  \GD p_i=p_i\lr{w_i-1}=p_i\Ga_i,
\end{equation*}
for which Fisher defined $\Ga_i=w_i-1$ as the average excess in fitness. Then the first term of the finite difference of the regression in \Eq{wD} is
\begin{equation}\label{eq:ftns}
  \GD\wbar_{\mathrm{ns}}=\bb\cdot\GD\bp = \sum_i b_i\GD p_i=\sum_i p_i\Ga_i b_i,
\end{equation}
which is often called the additive genetic variance in fitness, as shown in the following subsection. Fisher called this the fundamental theorem of natural selection. The result is that the partial change in mean fitness caused by the change in gene frequencies while holding constant the average effects is the additive genetic variance in fitness.

In Fisher's causal perspective, changes in gene frequencies, $\GD\bp$, are caused directly by natural selection, and changes in the partial regression coefficients, $\GD\bb$, are caused by change in context.

\subsection{Details on genetic variance}

This subsection shows why the expression on the right side of \Eq{ftns} is a genetic variance. These details can be skipped on first reading.

From \Eq{regZ}, we can write the regression expression for fitness as
\begin{equation*}
  w_j=\sum_i b_i x_{ij} + \Ge_j = g_j+\Ge_j,
\end{equation*}
in which $g_j=\sum_i b_i x_{ij}$ is the genetic value of an individual, and here we are indexing fitness by $j$ rather than $i$ as above. Thus, compactly, $w=g+\Ge$, the sum of the modeled genetic and residual effects. Next, recall that $q'_j=q_jw_j$ for our assumed mean fitness of $\wbar=1$, and $p_i = \sum_j q_j x_{ij}$, so that
\begin{align*}
  \GD p_i = \sum_j q'_j x_{ij} - \sum_j q_j x_{ij} &= \sum_j q_j\lr{w_j-1}x_{ij}\\
			&= \mathrm{Cov}(x_i,w).
\end{align*}
Then, in \Eq{ftns}, we can equivalently write
\begin{equation*}
  \sum_i b_i\GD p_i = \mathrm{Cov}\lr{\sum b_ix_i,w}=\mathrm{Cov}(g,w)=\mathrm{Var}(g),
\end{equation*}
because $g$ and $\Ge$ are uncorrelated by least squares regression \autocite{rao73linear,searle71linear,seber03linear}. Thus, the expression in \Eq{ftns} is the genetic variance for the particular model of genetic effects.

\subsection{Interpretation}

Fisher's theorem states, from the first term of \Eq{wD}, that the partial change in mean fitness caused by natural selection is $\bb\cdot\GD\bp$, freezing the partial regression coefficients $\bb$ at their initial values while letting gene frequencies $\GD\bp$ change. Everything that changes the mapping from the given genic predictors to fitness is pushed into the second term, $\bp'\cdot\GD\bb$.

It is obvious that Fisher's theorem does not fully account for how natural selection changes total mean fitness. The average effects of genes, which are the partial regression coefficients, typically depend on the gene frequencies that are changed by natural selection. So what does it mean to say that the direct cause of natural selection is only through changes in gene frequencies while holding constant the average effects of those genes?

There is also the problem that interactions between genes, such as dominance and epistasis, influence fitness. Those interactions depend on genotypic arrangements, which are altered by natural selection.

What was Fisher thinking? No one knows for sure. Certainly, Fisher understood all of this. He essentially invented the theory of classical statistical inference \autocite{fisher25theory,fisher22on-the-mathematical}. And his 1918 article was the first to express clearly the various components of variance in the expression of traits, including the components associated with dominance and epistasis \autocite{fisher18the-correlation}.

Our strongest clues about Fisher's thoughts come from his own words \autocite{fisher30the-genetical}. The first sentences of his great book on natural selection are: ``Natural Selection is not Evolution. Yet, ever since the two words have been in common use, the theory of Natural Selection has been employed as a convenient abbreviation for the theory of Evolution by means of Natural Selection, put forward by Darwin and Wallace. This has had the unfortunate consequence that the theory of Natural Selection itself has scarcely ever, if ever, received separate consideration.'' Put another way, Fisher's goal was absolutely focused on separating natural selection as a specific force from its wider context of evolutionary change.

Further, Fisher emphatically stated in 1941 that he had never written about the evolution of mean fitness, and wondered why Sewall Wright kept making the mistake of linking his fundamental theorem of natural selection to an overall conclusion about mean fitness \autocite{fisher41average}.

Fisher's way of isolating natural selection was to focus on gene frequency changes while holding constant the average effects of the genes. That does not work if one's goal is to calculate the dynamics of mean fitness evolution, as noted at the start of this subsection. But there is a certain logic to it that fits the common theories of causal analysis that developed subsequent to Fisher's writings \autocite{pearl09causality}.

In particular, Fisher noted that natural selection changes the frequencies of genes. That change in gene frequencies is directly transmitted to the next generation. Natural selection also changes genotype frequencies. But in standard Mendelian genetics, reassortment, recombination, and mating all change genotype frequencies while holding constant gene frequencies. So the pure direct effect of natural selection is only in the gene frequency changes. All else is context. 

\reviset{Thus, a natural causal counterfactual that one may use} to isolate natural selection from its broader evolutionary context is the change in gene frequency. That approach does not make the complexity of calculating the evolution of mean fitness go away. But it is consistent with the modern theory of causal analysis by counterfactual perturbation \autocite{pearl09causality}, which Fisher understood many decades earlier.

In particular, Fisher recognized that the partial regression coefficients (average effects) compute the counterfactual causal effect of randomized gene substitution, averaging over nonlinear genotypic interactions as they occur in the initial population, thereby isolating each gene's population-specific effect on fitness.

There is no resolution between those who find such attempts at isolating a force from its system as profoundly misleading and those who find such isolation as a good approach to partial causal understanding. My own view is that one should understand both perspectives.

The benefit of full exact calculation of total change is easy to understand. In favor of partial analysis, most learning and optimization methods work by partial perturbation, evaluating changes in one set of parameters while holding others fixed \autocite{nocedal06numerical,goodfellow16deep,frank25the-price}. One calculates partial perturbations because one can rarely do complete calculations for how changing a system alters its performance. For example, gradient descent in machine learning updates parameter weights by their partial derivatives with respect to a loss function, which is similar to Fisher's isolation of gene frequency changes while holding average effects constant.

Ultimately there are benefits in separating how one thinks about forces from how one analyzes dynamics. Lanczos emphasized this separation in his classic treatment of the variational principles in mechanics, in which he isolated individual forces from the full complexity of constrained motion \autocite{lanczos86the-variational}.

\reviset{
How closely does Fisher's approach match modern counterfactual causal analysis? In an earlier article \autocite{frank97the-price}, I showed that the Price equation combined with regression provides a common path-analytic causal framework that unifies Fisher's theorem, Robertson's \autocite{robertson66a-mathematical} covariance theorem for quantitative genetics, the Lande-Arnold \autocite{lande83the-measurement} model for the analysis of natural selection, and Hamilton's \autocite{hamilton70selfish} rule for kin selection.

Lee and Chow \autocite{lee13the-causal} later showed that Fisher's average effect corresponds to a causal intervention under Pearl's do-operator \autocite{pearl09causality}, a formal method for counterfactual analysis. Otsuka and Okasha extended a causal interpretation of the Price equation through Pearl's counterfactual methods \autocite{otsuka16causal,okasha20the-price}. Other studies similarly developed formal counterfactual analyses for selection gradients and kin selection \autocite{fromhage19the-strategic,henshaw20quantifying}.

These various studies develop increasingly elaborate formal machinery to arrive at the insight that Fisher's chain-rule partition can provide counterfactual causal attribution. The present article shows how simply this point arises from first principles.
}

\section{The Price equation}

The Price equation is another product rule expression for finite differences \autocite{price72extension,frank12naturalb}. This equation is widely used in ecology, evolutionary biology, and other disciplines. Define the population mean of a variable as $\zbar=\sum_j q_j z_j=\bq\cdot\bz$, in which $q_j$ is the frequency of the value $z_j$. Then, by our standard expression for finite differences
\begin{equation}\label{eq:priceEq}
  \GD\zbar=\GD\bq\cdot\bz \mskip5mu+\mskip5mu \bq'\cdot\GD\bz.
\end{equation}
\reviset{which is my generalization of the Price equation \autocite{frank97the-price,frank12naturalb}.} The Price equation \reviset{was originally} expressed in terms of covariance and expectation functions \autocite{price72extension}, which can be obtained by noting that we can write frequency changes in terms of fitness as in the prior section as $\GD q_j=q_j\lr{w_j-1}$, again assuming that $\wbar=1$ in the initial population. Also note that from the prior section we have for the relation between frequencies and fitness $q_j'=w_jq_j$. Then we can expand \Eq{priceEq} as
\begin{align*}
  \GD\zbar 	&=\sum_j q_j\lr{w_j-1}z_j+\sum_j q_jw_j\GD z_j\\
  			&=\mathrm{Cov}(w,z)+\mathrm{E}(w\GD z).
\end{align*}

\section{Partitions in other disciplines}

\subsection{Economics}

A simple example of the Oaxaca-Blinder decomposition \autocite{blinder73wage,oaxaca73male-female,fortin11decomposition} can be written for the difference in average income between two groups. For the first group, we start with the regression of income, $z$, on a vector of predictors, $\bx$, as
\begin{equation*}
  z_j = \bb\cdot\bx_j + \Ge_j
\end{equation*}
in which predictors can be variables such as education level. The equation follows the same assumptions as \Eq{regZ}. Because the average of the error term is zero, we have as before $\zbar=\bb\cdot\bxbar$. Then, we get the difference in mean incomes between groups as the same expression we found in \Eq{regressD} as
\begin{equation*}
  \GD\zbar=\bb\cdot\GD\bxbar \mskip5mu+\mskip5mu \bxbar'\cdot\GD\bb.
\end{equation*}
For example, suppose the first group has a particular relation between education and income, described by the regression coefficients, $\bb$. The second group has a different average for education level, $\GD\bxbar$. The first term predicts the change in income using the income-education relations in the first group. The second term describes how changes in the income-education relations in the second group alter the predicted change in income between the two groups.

Overall, we decompose the total change into the difference ascribed to changes in education holding constant the income-education relationship from the first group and the difference ascribed to changes in the income-education relationship holding the education levels constant from the values of the second group.

This decomposition was originally developed to study wage discrimination between groups defined, for example, by gender or race. The key question was how much of the observed wage gap could be attributed to differences in observable characteristics such as education, and how much might be attributed to differences in the relation between income and education, potentially reflecting discrimination.

The parallel with Fisher's theorem is direct but inverts emphasis. Fisher focused on the first term---the change in gene frequencies holding constant the regression coefficients---as the quantity of causal interest. Fisher's second term treats the shift in regression coefficients as a remainder reflecting changed context.

By contrast, economists often care most about the second term, because one cause for the shift in regression coefficients may be hidden discrimination, capturing the portion of the wage gap that cannot be explained by the differences in observable predictors. The different foci arise entirely from the causal question being asked.

\subsection{Demography}

Kitagawa decomposed differences between two demographic rates \autocite{kitagawa55components}. For example, the difference between the death rates of two populations can be partitioned into components. The first is the differences in age composition, holding age-specific mortality rates constant. The second is the differences in age-specific mortality rates, evaluated at the changed age composition.

This partition follows \Eq{regressD}, with fraction of individuals at each age playing the role of predictor means and age-specific mortality rates playing the role of regression coefficients. This decomposition is widely used in demography \autocite{fortin11decomposition}.

\subsection{Thermodynamics}

Nicholson et al.\ decomposed the change in the mean value of a thermodynamic ensemble into a term for changes in the probability distribution holding observable values constant and a term for changes in observable values holding the distribution constant \autocite{nicholson20timeinformation}. They used this decomposition to unify several results in irreversible thermodynamics and information theory.

Frank and Bruggeman showed that this thermodynamics decomposition is identical to the Price equation, given in \Eq{priceEq} \autocite{frank20the-fundamental}. All of these decompositions are applications of the simple product-rule partition in \Eq{finiteD}, which leads directly to the regression partitioning commonly used in applications.

\section{Discussion}

The product-rule partition and its application to regression are purely algebraic and hold without any causal assumptions. Causality enters only when one interprets the regression coefficients as effects, such as Fisher's average effects, and when one chooses a what-if counterfactual comparison that specifies what is held fixed across contexts \autocite{holland86statistics,pearl09causality}.

Fisher's theorem corresponds to the counterfactual in which the regression coefficients linking individual genes to fitness are held constant while gene frequencies change. The remainder term collects changes in that mapping due to genetic interactions, genetic background, and environment. This perspective explains both the usefulness of Fisher's partition for a particular causal question and why disagreements persist when readers prefer a different counterfactual analysis.

The same partition supports different causal questions depending on which term one treats as the quantity of interest. Fisher focused on the first term to isolate the direct force of natural selection. Economists often focus on the second term, in which shifts in regression coefficients may reveal discrimination. Demographers use both terms symmetrically to separate compositional change from rate change. In each case the mathematics is identical. What differs is the substantive question that motivates the partition.

\reviset{Fisher, the economists who developed the Oaxaca-Blinder decomposition, and the machine learning engineers who use back propagation all apply the same chain-rule mathematics. They differ only in the causal question they ask. Decades of econometric work on the Oaxaca-Blinder decomposition have addressed the choice of reference group, the sensitivity of the partition to the specific predictors used, and the conditions under which the decomposition supports causal inference \autocite{fortin11decomposition}. Those lessons apply to any chain-rule causal decomposition, including Fisher's theorem.
 
The chain rule is an elementary mathematical identity. The main point of this article is that much of causal analysis across the sciences reduces to this identity. Fisher saw it in 1930 and built his fundamental theorem of natural selection on it. The economists rediscovered it independently. Back propagation exploits the same principle to train the large models on which modern artificial intelligence depends. Recognizing the common foundation clarifies understanding, connects previously isolated disciplines, and reveals the deep simplicity beneath seemingly complex methods of causal attribution.
}

\section*{Acknowledgments}

\noindent I thank Jeff Zabel for pointing out the similarity between the Oaxaca-Blinder decomposition and some classic equations in biology, and for his general insights about causality across disciplines.

\section*{Data and code availability}

This work did not generate new data or code.

\section*{Funding}

The Donald Bren Foundation and National Science Foundation grant DEB--2325755 support my research.

\section*{Conflict of interest}

The author declares no conflict of interest.

%\vfill\eject

\mybiblio	% uses main.bib by default, add other bibs as needed

% used cuted package strip env to force balancing of columns
\ifmulticol\begin{strip}\hbox{\null}\end{strip}\hbox{\null}\fi

\end{document}